\documentclass[12pt]{iopart}
\usepackage{iopams}           
\usepackage{graphicx}         
\usepackage{bm}               
\usepackage{lineno}           

\usepackage[usenames]{color}               
\definecolor{MyDarkRed}{rgb}{0.30,0.08,0.0} 
\definecolor{MyDarkGreen}{rgb}{0.0,0.30,0.08} 
\definecolor{MyDarkBlue}{rgb}{0.0,0.08,0.30} 

\usepackage[
    bookmarks=true,         
    unicode=true,          
    pdftoolbar=true,        
    pdfmenubar=true,        
    pdffitwindow=true,      
    pdftitle={A Paramaterization of Cosmic Ray Shower Profiles Based on Shower Width},  
    pdfauthor={J.A.J. Matthews},     
    pdfsubject={Astroparticle Physics},   
    pdfcreator={J.A.J. Matthews},   
    pdfproducer={J.A.J. Matthews}, 
    pdfkeywords={cosmic-ray shower profile}, 
    colorlinks=true,        
    linkcolor=MyDarkBlue,          
    citecolor=MyDarkRed,        
    filecolor=MyDarkGreen,      
    urlcolor=MyDarkGreen         
]{hyperref}
\usepackage[all]{hypcap}    

\begin{document}

\newcommand{\secref}[1]{\sref{#1}}
\newcommand{\figref}[1]{\Fref{#1}}
\newcommand{\equref}[1]{\eref{#1}}
\newcommand{\tabref}[1]{Table \ref{#1}}
\newcommand{\degree}{\ensuremath{^{\circ}}}
\newcommand{\Nskies}{{10^{4}}}
\newcommand{\SP}{{\Sigma_{P}}}
\newcommand{\tal}{{\alpha}}
\newcommand{\talo}{{\alpha_{01}}}
\newcommand{\taloo}{{\alpha_{001}}}
\newcommand{\tbl}{{\beta}}
\newcommand{\tblo}{{\beta_{\talo}}}
\newcommand{\tbloo}{{\beta_{\taloo}}}
\newcommand{\ho}{{\mathsf{H}_{iso}}}
\newcommand{\hs}{{\mathsf{H}_{sig}}}
\newcommand{\nobs}{{n_{obs}}}
\newcommand{\nexp}{{n_{exp}}}
\newcommand{\lnPi}{{\ln P_{i}(\nobs | \nexp)}}
\newcommand{\lnPt}{{\ln P_{\theta}(\nobs | \nexp)}}
\newcommand{\lnPss}{{\ln P_{\gamma \zeta}(\nobs | \nexp)}}
\newcommand{\arXiv}[2]{\href{#1}{{\tt #2}}}

\title[A Parametrization of Cosmic Ray Shower Profiles Based on Shower Width]{A Parametrization of Cosmic Ray Shower Profiles Based on Shower Width}
\author{J.A.J.~Matthews, R.~Mesler, B.~R.~Becker, M.~S.~Gold, J.~D.~Hague }
\address{
  University of New Mexico, Department of Physics and Astronomy \\
  Albuquerque, New Mexico, USA
}
\ead{johnm@phys.unm.edu}

\begin{abstract}
Cosmic ray (CR) air showers, detected via the air fluorescence technique, are
reconstructed in part using functions that parameterize the longitudinal
profile of each shower.   The profile parameterization yields the position 
of shower maximum, $X_{max}$ in gm/cm$^2$, 
which is sensitive to the incident CR particle 
type: {\it e.g.} p, C/N/O, Fe or $\gamma$.  The integral of the profile 
is directly related to the shower energy.  The Pierre Auger fluorescence
reconstruction uses the Gaisser-Hillas (GH) 4-parameter form\cite{GH}.  The 
HiRes group has used both the Gaisser-Hillas form and a 3-parameter 
Gaussian in Age form\cite{GIA}.  
Historically analytic shower theory suggested yet other forms; the best
known is a 3-parameter form popularized by Greisen\cite{Greisen}.  Our work
now uses the shower full width half-maximum, $fwhm$, and shower 
asymmetry parameter, $f$, to
unify the parameterization of all three profile functions.  Furthermore
shower profiles expressed in terms of the new parameters: $fwhm, f$
have correlations greatly reduced over {\it e.g.} Gaisser-Hillas
parameters $X_0, \lambda$.   This allows shower 
profile reconstructions to add constraints (if needed) on the mostly 
uncorrelated parameters $fwhm$, $f$.
\end{abstract}

\vspace{2pc}
\noindent {\it Submitted to: J. Phys. G: Nucl. and Part. Phys.} \\
\noindent {\it Dated:} \today
\maketitle

\section{Introduction} \label{sec:intro}

Following the first observation of extensive air showers by
the air fluorescence technique\cite{bergeson}, the Fly's Eye experiment
pioneered analysis techniques\cite{baltrusaitis,cassiday,bird} that are 
largely unchanged to this day.  Most relevant to this paper are the 
analysis techniques needed to exploit the feature of fluorescence 
detectors to observe directly the development profile of 
air showers in the atmosphere.  The shower profile provides a
direct measurement of the depth of shower maximum, $X_{max}$ in
gm/cm$^2$, and provides 
an almost model-independent measurement of the cosmic ray energy\cite{bird}.
  
Historically, the shower profile was parameterized\cite{baltrusaitis} 
using three parameter Gaussian or four parameter Gaisser-Hillas\cite{GH} 
(GH) functions to 
characterize the shower brightness {\it versus} shower depth, $X$(gm/cm$^2$).
The Gaussian parameterization was appealing, as it depended on distance from
shower maximum, $X - X_{max}$, and depended on one parameter to characterize
the shower width.  As the precision of fluorescence detector
measurements improved the Gaussian parameterization was discarded, as 
it did not model the true shape of shower profiles that are asymmetric
in shower depth with respect to shower maximum.  The use of the 
four parameter GH function was often limited by the realities of data 
signal uncertainties or by the limited fraction of the shower profile 
observed for any given shower.  

More recently the HiRes/MIA collaboration\cite{GIA} analyzed a 
composite\footnote{Showers
observed by the HiRes prototype/MIA detectors were aligned to have a common
shower maximum.  The showers were then averaged to obtain a composite shower 
profile that was then used in their analysis.} 
shower profile using three
functional forms: Gaisser-Hillas\cite{GH} (noted previously), a form
motivated by analytic shower theory and popularized by Greisen\cite{Greisen} 
(Greisen), and a new form motivated by the observation that showers were
rather symmetric when plotted in shower age, termed {\it Gaussian in Age} 
(GIA)\cite{GIA}.  Interestingly, the three functions were almost 
indistinguishable and described the data comparably well.  The paper 
found a strong correlation between two of the GH parameters and 
suggested that a simpler (3 parameter) form
should be adequate to describe the shower profiles.
A closely related comparison of GH and GIA functions to
Monte Carlo simulated showers\cite{song} generated using CORSIKA\cite{heck},
again found comparably good agreement using GH or GIA profile functions.
Interestingly the Monte Carlo study\cite{song} observed the near equality 
of the width at half-maximum, $fwhm$, of proton, iron and photon showers 
but did not exploit this fact.

Motivated by the history above, in this paper we write the GH, GIA
and Greisen functions in terms of 4 parameters: the intensity at shower
maximum, $N_{max}$, the depth of shower maximum, $X_{max}$, and two
other parameters that in the case of GH are the parameters $X_0$ and
$\lambda$\footnote{While the conventional versions of GIA and Greisen 
are recovered setting $X_0 = 0$, the introduction of $X_0$ 
allows for a more symmetrical comparison to GH as well as allowing these 
forms to be applied to {\it e.g.} neutrino induced showers.}.
In all cases the profile functions can then be re-expressed in terms of
a physical distance $\Delta = X - X_{max}$, a composite parameter, 
$W \equiv X_{max} - X_0$, and a (forth) dimensionless parameter, $\xi$.
We will now show that $W$ depends both on shower width, {\it e.g.} $fwhm$,
and on a shower asymmetry parameter, $f$.  The parameter, $\xi$
depends only on the shower asymmetry, $f$.

\section{Profile functions} \label{sec:prof}

We now summarize the three profile functions: GH, GIA and Greisen.

\subsection{Gaisser-Hillas} \label{ssec:GH}
The Gaisser-Hillas function\cite{GH} is conventionally written:
$$ N(X)_{GH} ~=~ N_{max}~( { {X - X_0}\over{X_{max} - X_0} } )^{{X_{max} - X_0}\over{\lambda}}~e^{-{ {X - X_{max}}\over{\lambda}}} $$

\noindent If we introduce the new variable: $W \equiv X_{max} - X_0$ and
the distance from shower maximum: $\Delta = X - X_{max}$, the GH function
can then be written as:
\begin{equation}
N(X)_{GH} ~=~ N_{max}~( 1 + {{\Delta}\over{W}} )^{{W}\over{\lambda}}~e^{-{ {\Delta}\over{\lambda}}}
\end{equation}

\noindent It is useful to introduce two dimensionless 
quantities: $\epsilon \equiv {{\Delta}\over{W}}$
and $\xi \equiv {{W}\over{\lambda}}$.  Then a minimal form, useful
for comparisons, is obtained:

\begin{equation}
N(X; N_{max}, X_{max}, X_0, \lambda)_{GH} ~=~ N_{max}~( 1 + \epsilon )^{\xi}~e^{-(\epsilon ~{\xi} )}
\end{equation}

\noindent This form emphasizes the role of a physical quantity,
the distance from shower maximum, $\Delta$, in comparison to {\it e.g.} 
an unphysical quantity, $X - X_0$.  Furthermore $\Delta$ is scaled by a 
parameter $W$ potentially resolving a tension between $X_{max}$ and 
$X_0$ in a parameter optimization to {\it fit} experimental shower profiles.

While the new variables, $W,\xi$, are less intuitive than
the standard GH parameters $X_0,\lambda$, we will show that they are
related to two very intuitive parameters: the shower width, $fwhm$ and 
a dimensionless shower asymmetry parameter, $f$.

For later reference the integral of the GH function\cite{unger} is given by:

\begin{equation}
 \int^{\infty}_0 N(X)_{GH} dX = N_{max}~W~\xi^{-(\xi + 1)}~e^{\xi}~\Gamma(\xi + 1) 
\end{equation}

\noindent where $\Gamma$ is the Euler Gamma function.

\subsection{Gaussian in Age} \label{ssec:GIA}
The Gaussian in Age function\cite{GIA} is conventionally written:
$$ N(X)_{GIA} ~=~ N_{max}~e^{-{{1}\over{2}}~({{s - 1}\over{\sigma}})^2} $$

\noindent where ``$s$'' is the shower age and $s = 1$ at shower maximum.  
Motivated by the definition of the GH function, the
conventional definition for $s_{conv} = {{ 3 X}\over{X + X_{max}}}$ is 
then modified by replacing: $X \rightarrow X - X_0$ and 
$X_{max} \rightarrow X_{max} - X_0$.  Then with the same
definitions as used for the GH function (above), the redefined age 
variable is: $s(\epsilon) = {{1 + \epsilon}\over{1 + \epsilon/3}}$.  
Note that 
this definition preserves $s = 1$ at shower maximum and $s=3$ in the limit 
of $X>>X_{max}$,~but differs from ``$s_{conv}$''
at other shower depths.  As with the GH function, a minimal form for the GIA 
function is obtained:

\begin{equation}
N(X; N_{max}, X_{max}, X_0, \sigma)_{GIA} ~=~ N_{max}~e^{-2({{\epsilon}\over{(3 + \epsilon)\sigma}})^2}
\end{equation}

\subsection{Greisen} \label{ssec:Greisen}
The Greisen function\cite{Greisen} is conventionally written\cite{Greisen,GIA}:
$$ N(t)_{Greisen} ~=~ N_{max}~e^{(~t(1 -{{3}\over{2}}ln(s))~-~t_{max}~)} $$

\noindent where $t$ is the depth in the shower in radiation lengths,
$t_{max}$ is the depth of shower maximum and ``$s$'' is shower age.  
Following the modification of
GIA: $t \rightarrow t - t_0$ and $t_{max} \rightarrow t_{max} - t_0$, then
converting from depth in radiation lengths to depth in gm/cm$^2$,
$t = {{X}\over{36.7}}$, and using the same notation as for GH and GIA (above) 
we obtain the minimal form:

\begin{equation}
N(X; N_{max}, X_{max}, X_0, p_{36.7})_{Greisen} ~=~ N_{max}~e^{(~\epsilon(1 -{{3}\over{2}}ln(s(\epsilon)))~-~{{3}\over{2}}ln(s(\epsilon))~)~{{W}\over{p_{36.7}}}}~~~~
\end{equation}

\noindent where the conversion constant ``36.7gm/cm$^{2}$'' 
(per radiation length) is treated as a parameter\cite{GIA}, 
$p_{36.7}$\footnote{The original Greisen functional form is simply 
regained by setting: $X_0 = 0$ and $p_{36.7} = 36.7$.}.

It is interesting to note that Eqn.~2, 4 and 5 depend only on 
$\epsilon \equiv {{X - X{max}}\over{W}}$ and one additional dimensionless 
parameter: 
$\xi \equiv {{W}\over{\lambda}}$, $\sigma$ or ${{W}\over{p_{36.7}}}$.
As noted above, 
we will now show that $W$ depends on both shower width and asymmetry:
$fwhm$, $f$ and $\xi$ depends only on shower asymmetry: $f$.

\subsection{Profile parameters: $fwhm$ and $f$} \label{ssec:ProfParms}

While there are now some similarities between the three profile functions,
a simple connection between the profile parameters and the appearance of
a shower does not exist.  A natural choice is to use the width of the shower
(at {\it e.g.} half-height), $fwhm$ and an asymmetry parameter, $f$.  Using
the profile in Fig.~1(left), $fwhm \equiv \mathcal{L} + \mathcal{R}$ and 
$f \equiv \mathcal{L}/(\mathcal{L} + \mathcal{R})$.  Thus
$fwhm$ has dimensions (gm/cm$^2$) and $f$ is dimensionless.  Furthermore, 
because real and simulated
shower profiles rise more steeply than they fall $f$ will always be
less than 0.5.


\begin{figure}[h]
  \begin{tabular}{cc}
    \includegraphics[angle=-90.,width=7.5cm]{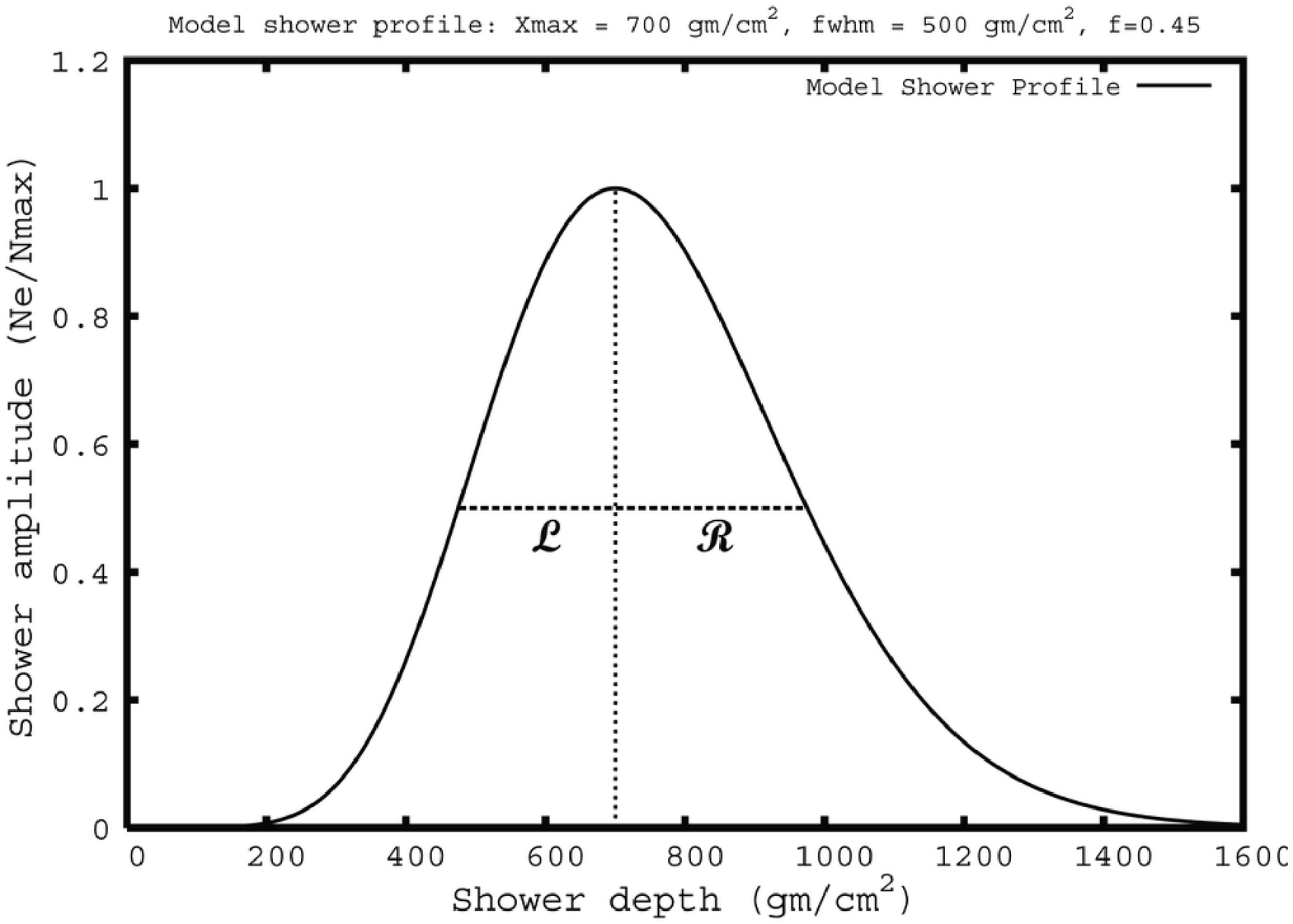} &
    \includegraphics[angle=-90.,width=7.5cm]{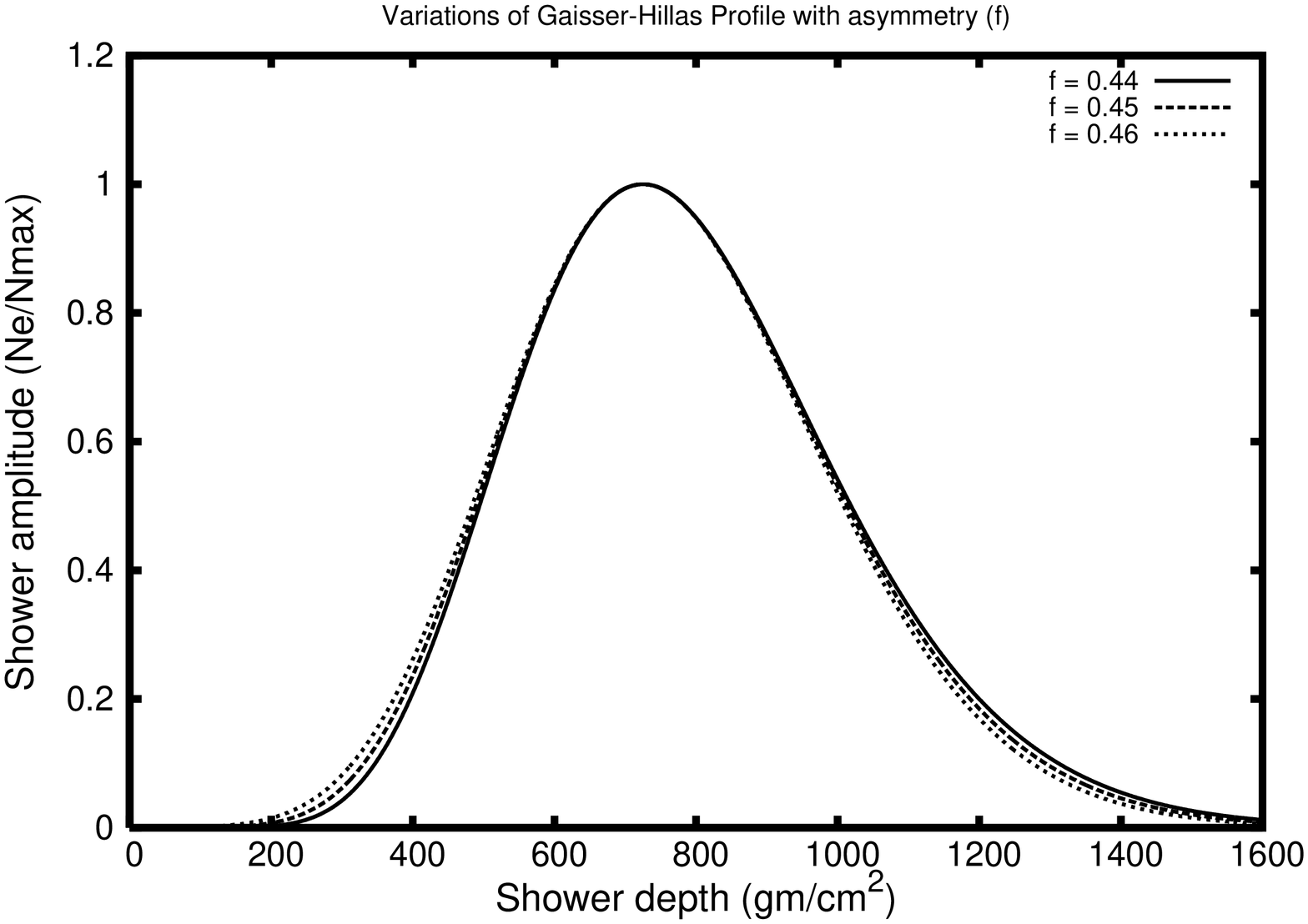}
  \end{tabular}
\caption{\label{fig:fig1}
{\bf Left:} Plot of a representative Gaisser Hillas (GH) shower profile
defined by: $X_{max} = 700$ gm/cm$^2$,
$fwhm \equiv \mathcal{L}+\mathcal{R} = 500$ gm/cm$^2$ and shower 
asymmetry $f \equiv \mathcal{L}/(\mathcal{L}+\mathcal{R}) = 0.45$. 
The vertical axis is the shower intensity ($N_e$) normalized to the intensity
at shower maximum ($N_{max}$) and the horizontal axis is the shower 
depth ($X$) in the atmosphere in gm/cm$^2$.
{\bf Right:} GH shower profiles with 
$X_{max} = 725$ gm/cm$^2$, $fwhm = 525$ gm/cm$^2$ 
and three different values of asymmetry: $f = 0.44$, 0.45 and 0.46.
The corresponding GH parameters are: ($X_0 = 2.2$ gm/cm$^2$, 
$\lambda = 68.3$ gm/cm$^2$), ($X_0 = -145$ gm/cm$^2$, 
$\lambda = 56.9$ gm/cm$^2$) and ($X_0 = -365$ gm/cm$^2$, 
$\lambda = 45.5$ gm/cm$^2$) respectively.
}
\end{figure}

CORSIKA simulations\cite{song} had suggested that shower $fwhm$ values were
rather similar for p, iron and $\gamma$ showers.  Results of CONEX\cite{conex,qgsjet,sibyll}
simulated showers are shown in Fig.~2 for $fwhm$ 
and for the asymmetry parameter, $f$,
for shower energies relevant to the Auger\cite{auger_FD}, HiRes\cite{hires} 
and TA\cite{TelArray} experiments.
These simulations are consistent with the CORSIKA results and suggest that
for energies $\sim 10^{18.5}$eV and for all primaries the $fwhm$ is in
the rather limited range:
475 gm/cm$^2$
 ~$^{<}_{\sim} ~fwhm~ ^{<}_{\sim} ~~575$ gm/cm$^2$
and asymmetry in the range: 0.44 ~$^{<}_{\sim} ~f~ ^{<}_{\sim} ~~0.47$.
Furthermore the CONEX simulations suggest that the asymmetry parameter, $f$,
may provide some discrimination in primary composition\cite{lisbon,viviana}, 
{\it i.e.} between different types of primary particles.  
Finally both $fwhm$ and asymmetry, $f$, show a slow, approximately
logarithmic increase with increasing shower energy.

\begin{figure}[h]
  \begin{tabular}{cc}
    \includegraphics[angle=-90.,width=7.5cm]{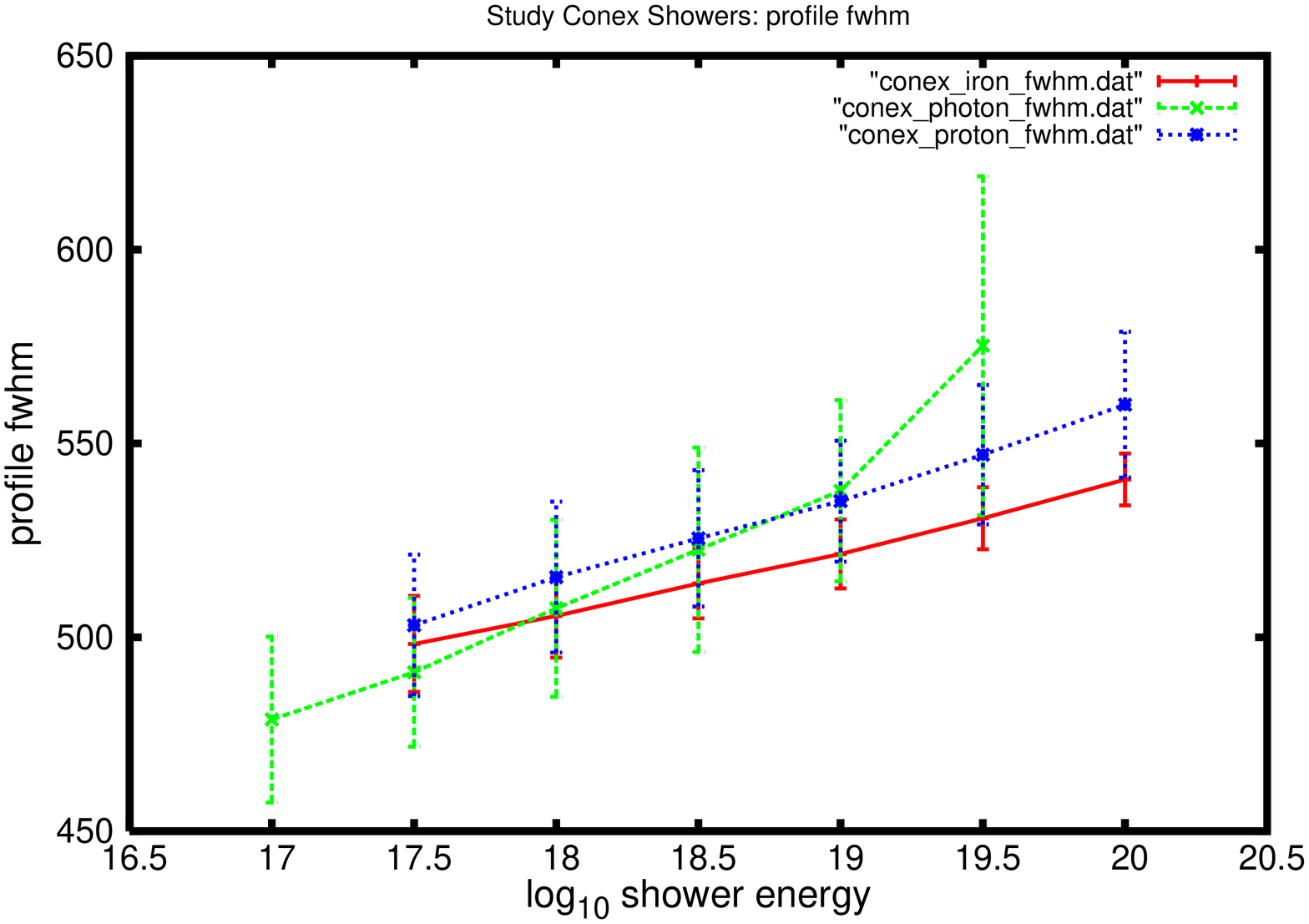} &
    \includegraphics[angle=-90.,width=7.5cm]{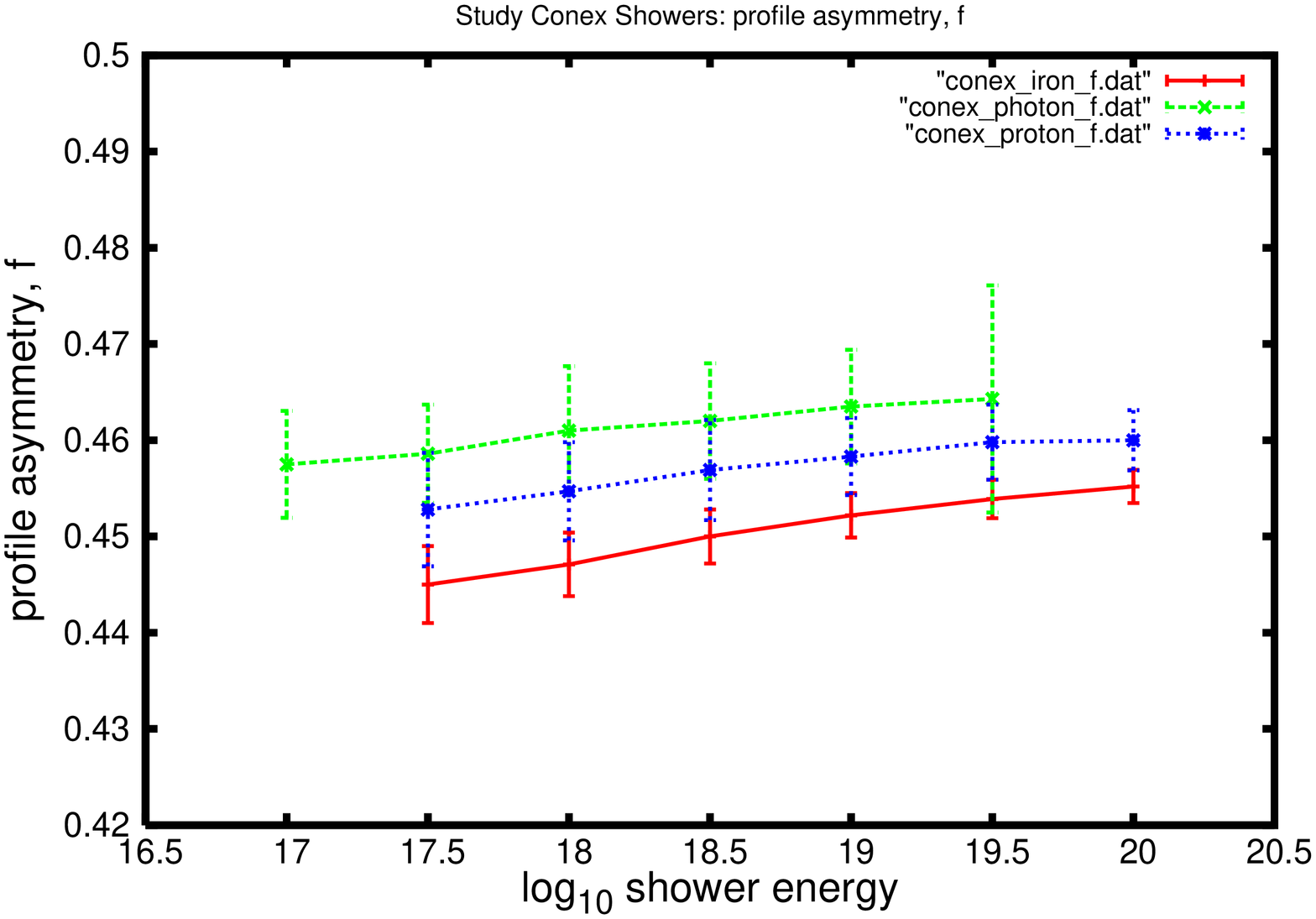}
  \end{tabular}
   \caption{\label{fig:fig2}
     The results of CONEX\cite{conex} simulations of
     proton, iron and $\gamma$ initiated air showers are shown in
     blue, red and green respectively.
      {\bf Left:} Plot showing the shower widths ($fwhm$)
     on the vertical axis
     {\it versus} the shower energies on the horizontal axis.
     The error bars give the RMS of the simulated
     showers in each energy bin.
     {\bf Right:} Plot showing the shower asymmetries ($f$) on 
     the vertical axis
     {\it versus} the shower energies on the horizontal axis.
}
\end{figure}

\subsection{Relating {\it old} and {\it new} shower parameters} \label{ssec:OldNew}

To relate the {\it old} to the {\it new} shower parameters, the shower
profile is evaluated at a fraction $h$ of maximum.
$N(X) = h ~ N_{max}$.   To illustrate the procedure we choose
the GH profile as written in Eqn.~1.  Then solving for the values of 
$\Delta$ that satisfy $N(X) = h ~ N_{max}$ one obtains:

$$ {{W}\over{\lambda}} ln( 1 + {{\Delta}\over{W}} ) ~-~ 
{{\Delta}\over{\lambda}} = ln(h) $$

\noindent This relation is true for two values of $\Delta$, see Fig. 1(left): 
~~$\Delta_{\mathcal{L}} = (-f(h)) ~ Width(h)$
and $\Delta_{\mathcal{R}} =  (1-f(h)) ~ Width(h)$ where 
$|\Delta_{\mathcal{L}}| + \Delta_{\mathcal{R}} = Width(h)$ is the
width of the shower profile curve at fractional {\it height} $h$\footnote{While
both $Width(h)$ and $f(h)$ depend on the choice for $h$, the shower shape and
the parameters describing the shower shape, {\it e.g.} GH parameters
$X_0, \lambda$, do not.}.  Following the familiar convention to describe
a profile at {\it half-height}, we define $fwhm \equiv Width(h=0.5)$ and 
$f \equiv f(h=0.5)$.

\begin{figure}[h]
  \begin{tabular}{cc}
    \includegraphics[angle=-90.,width=7.5cm]{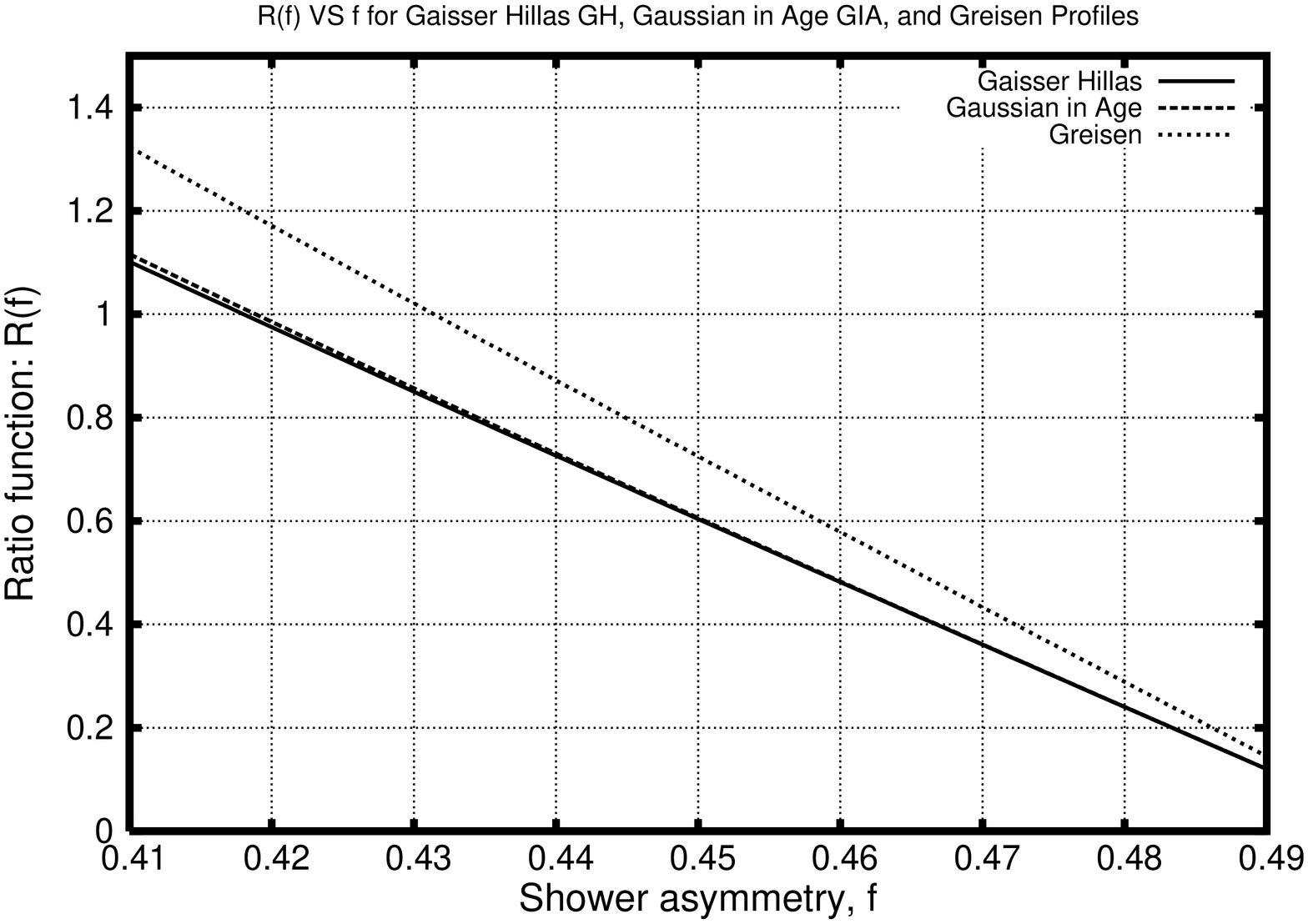} &
    \includegraphics[angle=-90.,width=7.5cm]{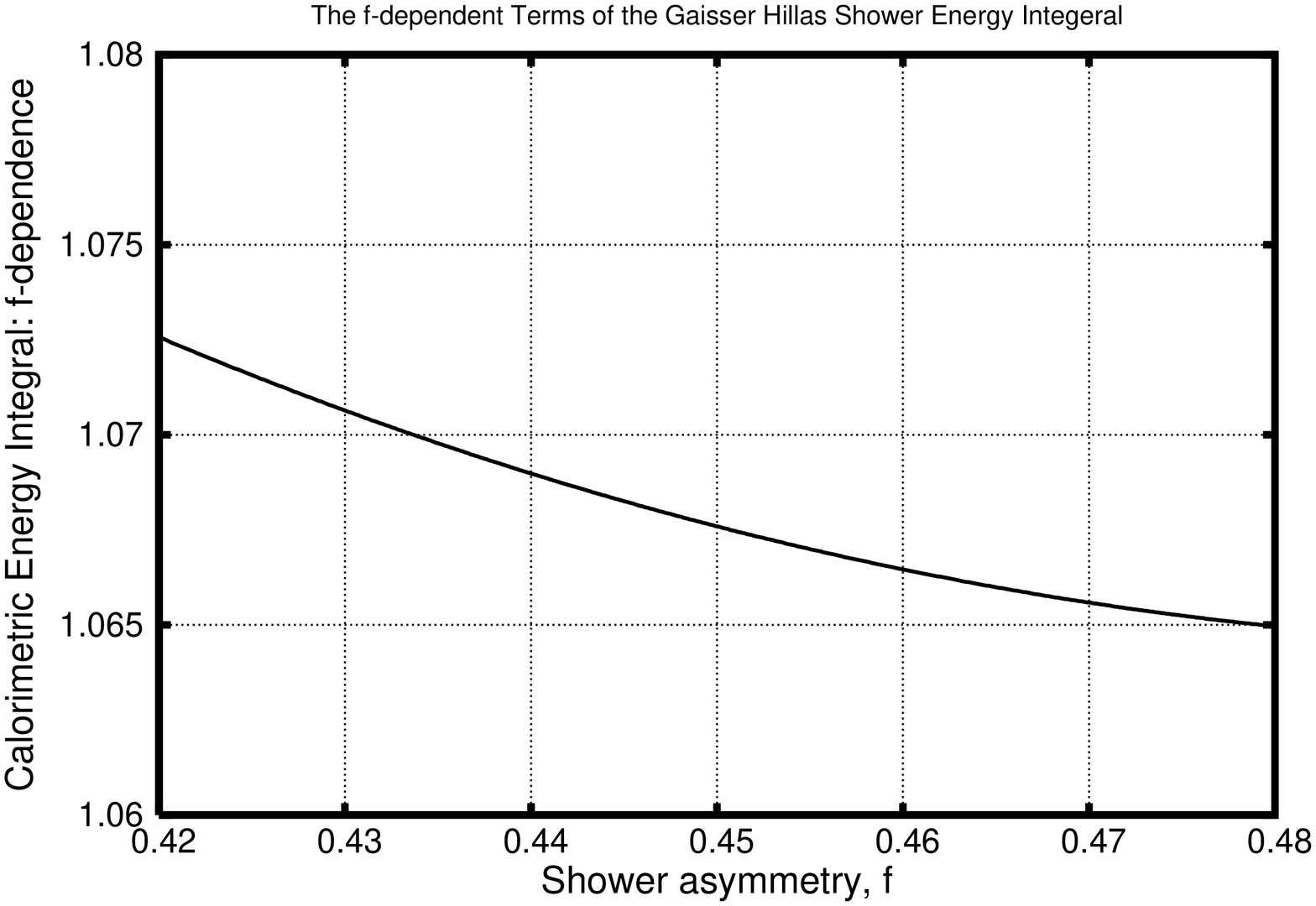}
  \end{tabular}
   \caption{\label{fig:fig3}
     {\bf Left:} The curves show the ratio function, 
     $R(f)$, plotted
     on the vertical axis {\it versus} the shower asymmetry, $f$,
     on the horizontal axis.  As discussed in the 
    text the ratio function:
 $R \equiv {{fwhm}\over{W}} = {{fwhm}\over{X_{max} - X_0}}$
 depends only on the shower asymmetry.  The ratio functions for the different
     shower profiles are plotted as: GH (solid), GIA (dashes) and 
     Greisen (dots).
     {\bf Right:} The asymmetry $f$-dependent terms in Eqn.~7,
     ${{\xi^{-(\xi + 1)}~e^{\xi}~\Gamma(\xi + 1)}\over{R(f)}}$, are
     plotted on the vertical axis {\it versus} the shower asymmetry, $f$,
     on the horizontal axis.
}
\end{figure}

The equations for $\Delta_{\mathcal{L}}$ and $\Delta_{\mathcal{R}}$
can be combined to obtain one (GH profile specific) relationship:

\begin{equation}
   ln( {{1 - f R}\over{1 + (1-f)R}} ) + R = 0
\end{equation}

\noindent where $R \equiv {{fwhm}\over{W}}$ depends only on the shower
asymmetry, $f$. 
Eqn.~6 is solved numerically to obtain the ratio function: $R = R(f)$; 
the result is shown in Fig.~3(left).  

The solution to Eqn.~6 relates $W$ to the 
shower $fwhm$: ~$W = {{fwhm}\over{R(f)}}$.  Finally with the ratio function
$R(f)$ and $W$ known, the GH parameters $X_0$ and $\lambda$ are given by:

$$ X_0 ~=~ X_{max} - W $$

$$ \lambda ~=~ {{ W ~ [~ln(1 - fR) ~+~ fR~] }\over{ln(h=0.5)}}  $$

It is important to note that $\xi = {{W}\over{\lambda}}$ depends only
on the asymmetry parameter, $f$.  Thus the calorimetric shower energy,
given by:

$$ E^{calor}_{shower} ~=~ \int^{\infty}_0 N(X)_{GH}~dE/dx~ dX $$

\noindent is to a good approximation\cite{songdEdx}, see Eqn.~3:

\begin{equation}
E^{calor}_{shower} ~=~ <dE/dx>~N_{max}~fwhm~({{\xi^{-(\xi + 1)}~e^{\xi}~\Gamma(\xi + 1)}\over{R(f)}})
\end{equation}

\noindent where $<dE/dx>~\approx~2.19$ MeV/particle/(gm/cm$^2$)\cite{songdEdx}
and $N_{max}$ is the number of charged particles at shower maximum.  The
dependence on shower asymmetry, $f$ is shown in Fig.~3(right) and is
$\sim 1$\% over the range of asymmetries predicted using CONEX, 
see Fig.~2(right).  Thus to a good approximation the
shower calorimetric energy  is proportional to: $N_{max} ~ fwhm$;
see Fig.~4.  
This relation emphasizes the importance of the shower width, $fwhm$, as
errors in determining the width directly propagate into errors in
estimating the shower energy.
This relation also shows that mis-measuring the asymmetry has almost no
effect on the shower energy.  Eqn.~7 and Fig.~3(right) do imply a 
small correlation coming from the calorimetric shower energy constraint:
namely that more asymmetric
showers (smaller values of $f$) should have slightly smaller values of $fwhm$.

\begin{figure}[h]
\centerline{  \includegraphics[angle=-0,width=12cm]{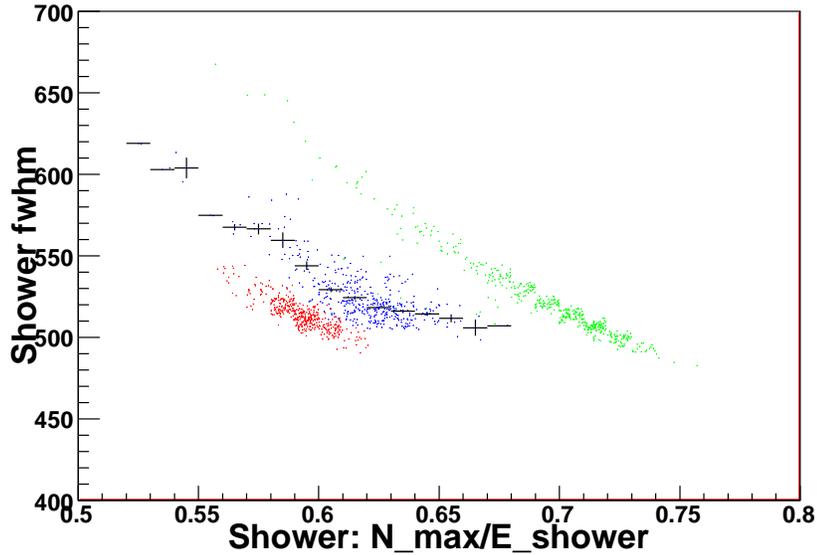} }
   \caption{\label{fig:fig4} 
    Scatter plot showing CONEX simulations of proton, iron and $\gamma$
  initiated air showers shown in blue, red and green respectively.  The points
    report the shower $fwhm$ in (gm/cm$^2$) on the vertical axis
    {\it versus} the ratio of shower quantities: $N_{max}/E_{shower}$ in
    (particles/GeV) on the horizontal axis.  See Eqn.~7 for motivation.
    All showers had an
    energy near $10^{18.5}$eV.  The relative
    (horizontal) displacement of the iron ~:~p~:~$\gamma$ showers is related
    to increasing missing energy (in muons and neutrinos) in 
    iron~{\it versus}~p~{\it versus}~$\gamma$ showers. 
    The points with error-bars reflect
    the average of the proton points in the scatter plot.
}
\end{figure}

To relate conventional GIA parameters, $X_0, \sigma$, to new parameters,
$fwhm, f$, we follow the same procedure that was employed for
the GH function.  For the GIA profile, the simple 
mathematical form of the GIA function 
(see Eqn.~4) results in an analytic (GIA profile specific) solution for 
the ratio function $R(f)$:

          $$  R(f) ~=~ {{3 (1 - 2f)}\over{2 f (1 - f)}} $$

\noindent which is plotted in Fig.~3(left).
As with GH,  $W~=~{{fwhm}\over{R(f)}}$.  The conventional GIA
parameters are then given by:

                 $$ X_0 ~=~ X_{max} - W $$

$$ \sigma ~=~ { {fR}\over{ (3 - fR) ~ \sqrt{ {{-ln(h=0,5)}\over{2}} } }} $$

Finally to relate conventional Greisen parameters, $X_0, p_{36.7}$, 
to new parameters, $fwhm, f$, the first step is to solve for the
values of $\Delta$ that satisfy $N(X) = h ~ N_{max}$:

 $$ {{\Delta}\over{W}} ~ ( 1 - {{3}\over{2}}{ln(s)} ) 
    ~-~ {{3}\over{2}}{ln(s)}
     ~=~ {{p_{36.7}}\over{W}}ln(h) $$

\noindent As with the GH function, this relation is true for two values of
$\Delta$: $\Delta_{\mathcal{L}}$ and $\Delta_{\mathcal{R}}$.
The two resulting equations can be combined into one relationship
that relates $R \equiv {{fwhm}\over{W}}$ to the shower asymmetry, $f$.
The solution, found numerically, is plotted in Fig.~3(left).  Once
the ratio function $R(f)$ is determined, then the Greisen 
parameters, $X_0, p_{36.7}$ are given by:

            $$ X_0 ~=~ X_{max} - W $$

   $$ p_{36.7} ~=~ {{W}\over{ln(h=0.5)}} ~ [ - f R ( 1 - {{3}\over{2}} ln(s_{\mathcal{L}})) ~-~  {{3}\over{2}}{ln(s_{\mathcal{L}})} ]  $$

\noindent where $s_{\mathcal{L}} \equiv s( {{\Delta_{\mathcal{L}}}\over{W}} )$ is the (redefined)
shower age (see Sect. 2.2 above).

\begin{figure}[h]
  \begin{tabular}{cc}
    \includegraphics[angle=-90.,width=7.5cm]{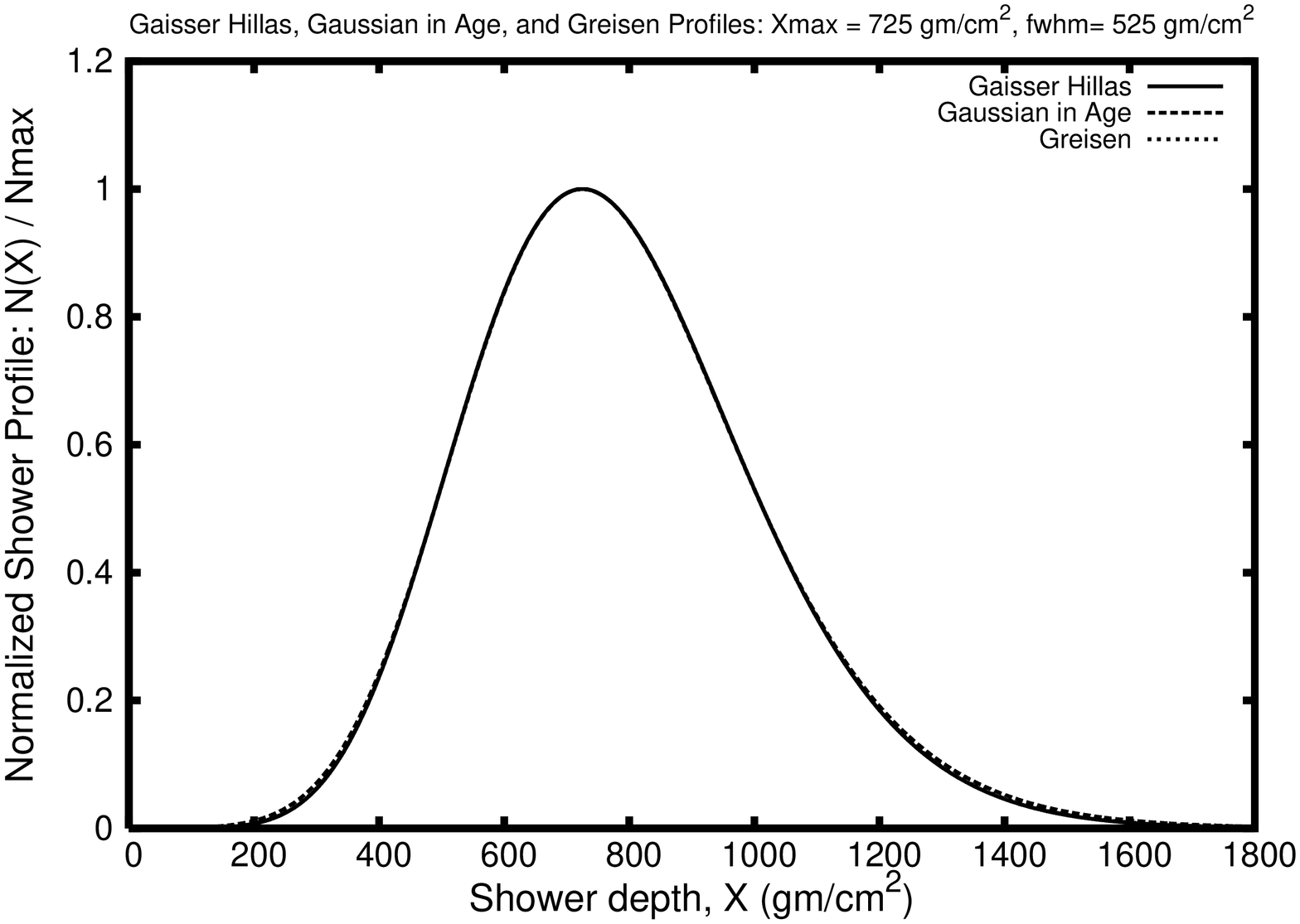} &
    \includegraphics[angle=-90.,width=7.5cm]{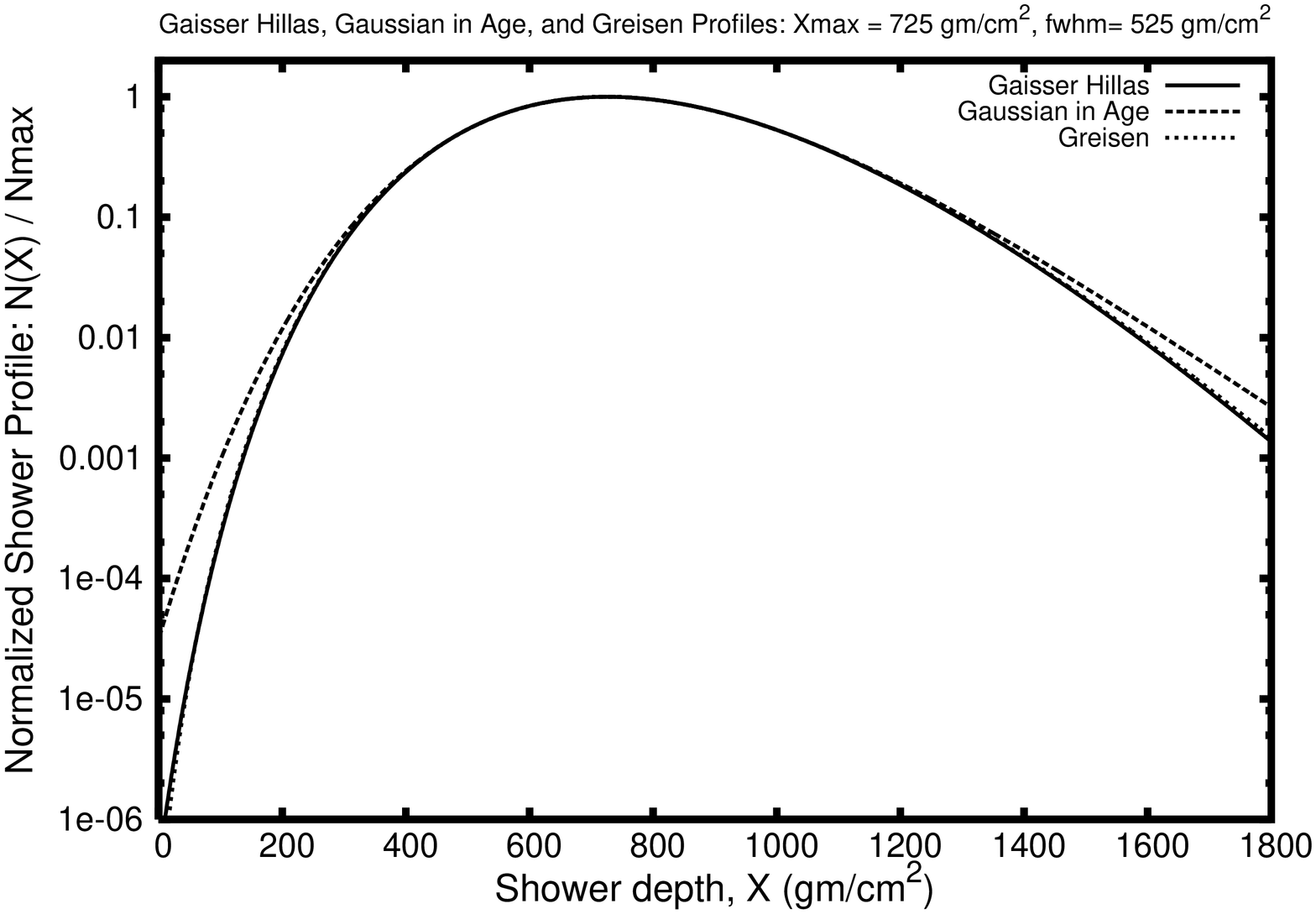}
  \end{tabular}
   \caption{\label{fig:fig5}
     {\bf Left:} Plot of shower profiles for GH (solid), GIA (dashes) 
     and Greisen (dots) profile functions for typical shower profile
     parameters:
     $X_{max} = 725$ gm/cm$^2$, $fwhm = 525$ gm/cm$^2$, and $f = 0.45$.
     The vertical axis is the shower intensity normalized 
     to the intensity at shower maximum and the horizontal axis is the shower 
     depth in the atmosphere in gm/cm$^2$. 
     {\bf Right:} Same shower profiles now with a semi-log plot to 
     show the small differences between the functions.
}
\end{figure}

\section{Profile comparisons} \label{sec:prof_comp}

Now that the GH, GIA and Greisen functions
can be expressed in common parameters: 
$fwhm, f$, it is easy and instructive to compare their shower profiles.
An example comparison, shown in Fig.~5, finds that GH and Greisen functions
are almost indistinguishable while (both) being very similar to, but
systematically below the GIA profile for shower depths well away from
shower maximum.  This result is consistent with HiRes/MIA\cite{GIA} and 
HiRes\cite{song} observations that the profiles describe experimental
(and simulated) shower
data comparably well near shower maximum.  Quantitatively,
the small but systematic profile function differences result 
in shower calorimetric energies evaluated using the GIA function being 
$\sim 1$\% larger than those evaluated using GH or Greisen forms.

As noted in Sect 2.4, the shower profiles away from shower maximum 
may provide some discrimination on cosmic ray composition\cite{lisbon,viviana} 
(cf. Fig.~2(right))
and on whether the GIA\cite{giller} 
{\it versus} GH (or Greisen) profiles provide 
somewhat better parameterizations of actual shower profiles.  
While interesting,
this is not the thrust of this paper and will not be discussed further.

\begin{figure}[h]
  \begin{tabular}{cc}
    \includegraphics[angle=-0,width=7.5cm]{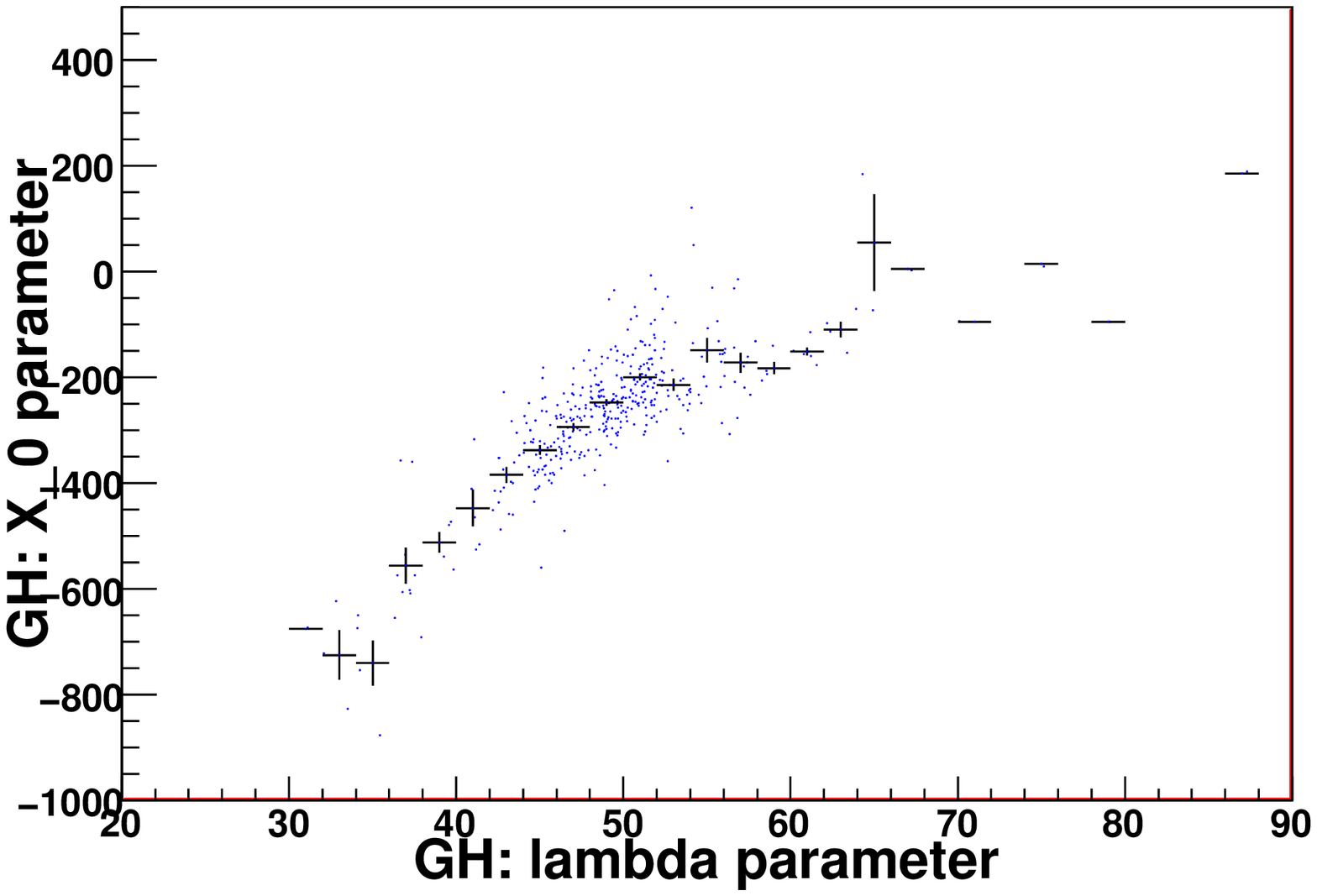} &
    \includegraphics[angle=-0,width=7.5cm]{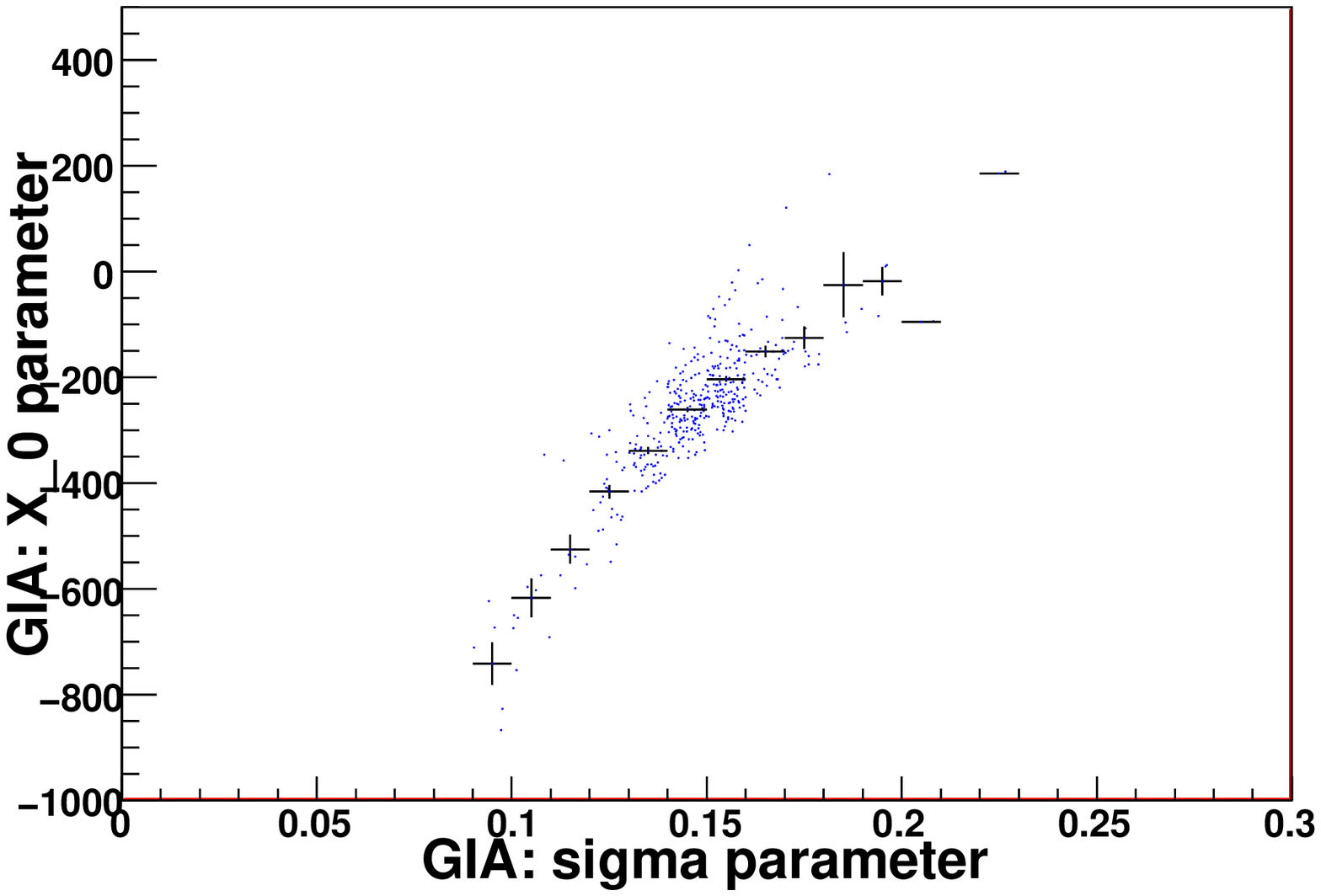} \\
    \includegraphics[angle=-0,width=7.5cm]{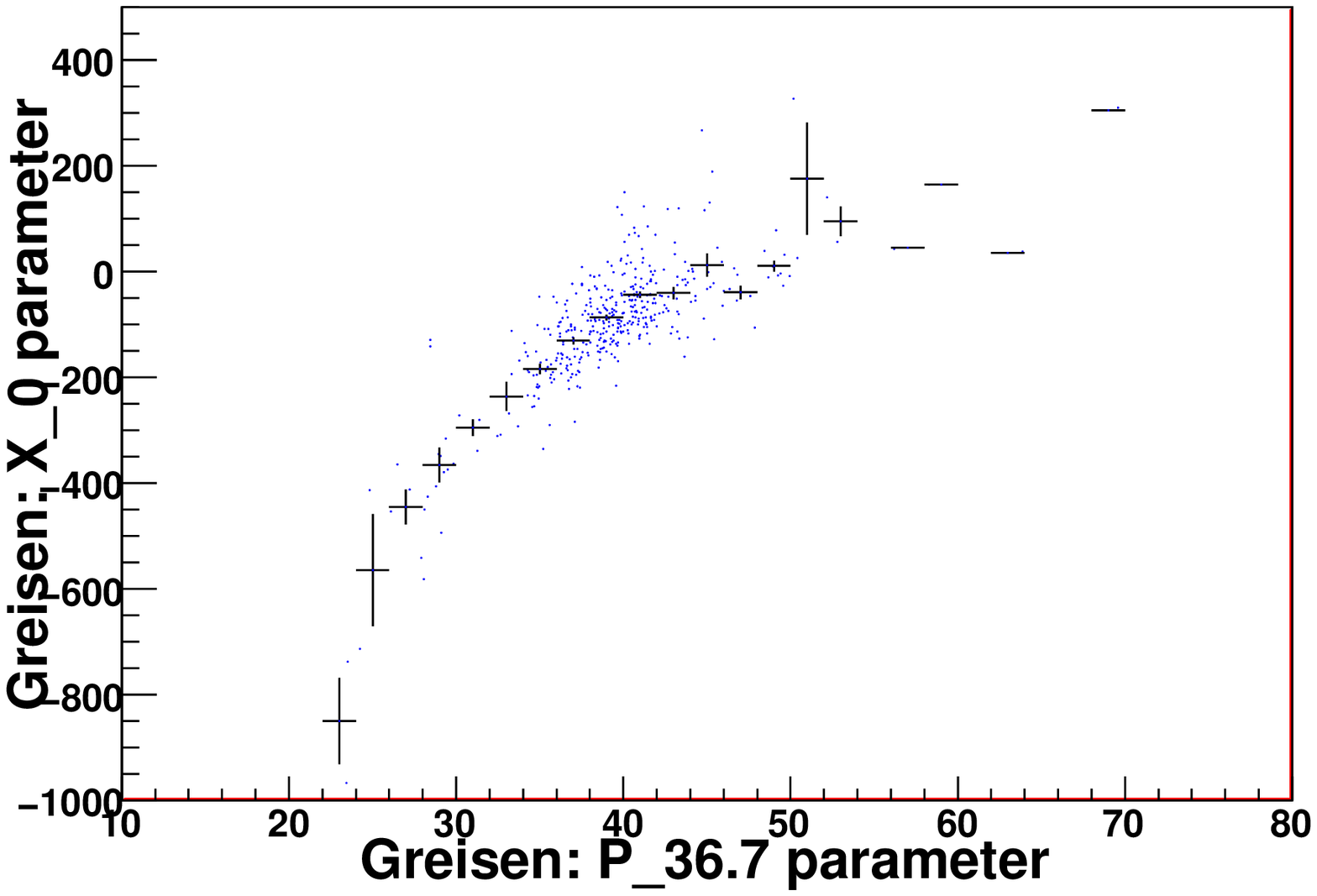} &
    \includegraphics[angle=-0,width=7.5cm]{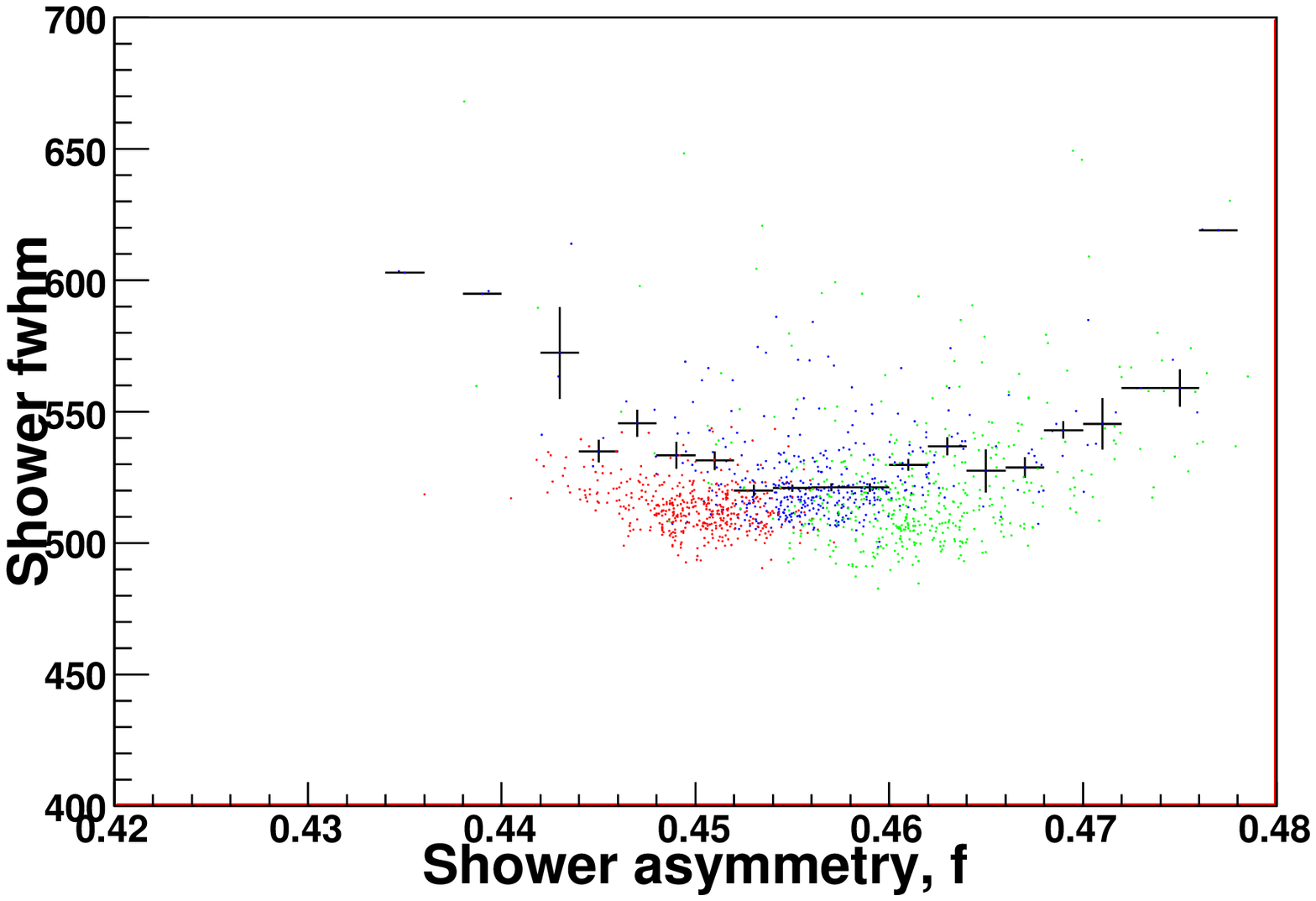} 
  \end{tabular}
   \caption{\label{fig:fig6}
     Scatter plots of {\bf Top Left:} GH parameters ($X_0$, $\lambda$),
     {\bf Top Right:} GIA parameters ($X_0$, $\sigma$),
     {\bf Bottom Left:} Greisen parameters ($X_0$, $p_{36.7}$) for
     proton and
     {\bf Bottom Right:} shower ($fwhm$, $f$) for proton (blue), 
     iron (red) and $\gamma$ (green) showers near
     $10^{18.5}$eV simulated with CONEX.  
     In all cases the first parameter is plotted on the vertical
     axis and the second parameter on the horizontal axis.
     The points with error-bars reflect
     the average of the (proton) points in the scatter plot.  The plot
     statistics
     are: 411 proton, 440 iron and 462 photon CONEX showers.
}
\end{figure}

A remaining issue is how best to apply four parameter shower profile
functions to less than ideal data.  In these circumstances the tendency
is to fix, or constrain, one or more of the parameters. An equivalent
point of view is the assumption that $X_0 = 0$ gm/cm$^2$ 
in the GIA or Greisen function
and/or that the parameter $p_{36.7} = 36.7$ gm/cm$^2$ (per radiation length)
in the Greisen function.

To study this issue, shower profile parameters from CONEX showers are 
shown in Fig.~6 for proton showers with energies $\sim 10^{18.5}$eV.
For each simulated shower, the shower $fwhm$ and asymmetry, $f$, are
determined.  The values are plotted in Fig.~6(lower right).  
Following Sect. 2.5 the $fwhm$ and $f$ values are used to determine
the corresponding parameters for GH, GIA and Greisen profiles.  The
results are plotted in the other quadrants of Fig.~6.  The conventional
shower shape parameters: ($X_0$,$\lambda$), ($X_0$,$\sigma$), and
($X_0$,$p_{36.7}$) show significant correlations while ($fwhm$, $f$) do not. 
In particular the correlation coefficients for GH, GIA and Greisen
parameters are in the range: $0.76 \sim 0.91$ 
for CONEX\cite{conex,qgsjet,sibyll}
proton and iron showers while the correlation coefficients for 
($fwhm$, $f$) are $\sim 0.12$
for proton showers and $\sim -0.11$ for iron showers.   

While the GH correlation results appear
consistent with the observation of parameter correlations in 
Ref.\cite{GIA}, our analysis does not (then) support the
use of GIA (or Greisen) shower profiles with the $X_0$ parameter set
to zero\cite{GIA}.  Rather than the correlation being evidence that
three parameters are sufficient to describe shower profiles,
it is more likely that GH, GIA and Greisen parameters
need to fulfill the calorimetric energy constraint of Eqn.~7\footnote{An 
almost equivalent statement, is that {\it e.g.} ($X_0$,$\lambda$) must
be compatible with the full with half maximum of the shower.}.
As noted above, this constraint is almost
insignificant with the choice of variables ($fwhm$, $f$).  In summary
the ($fwhm$, $f$) shower shape parameters appear less 
correlated that the other shower parameter pairs (while still assuming
a wide range of values).

That said, it is likely that shower details 
lead to correlations with any choice of shower parameters.  
In that regard, this study
is compatible with shower reconstructions using partial 
constraints\cite{unger}; we recommend that the constraints be
applied to $fwhm$ and $f$ rather than the GH parameters $X_0$ and $\lambda$.

\section{Conclusion} \label{sec:conc}

This analysis has studied three functions used to parameterize the
shower profiles of ultra-high energy cosmic rays: Gaisser-Hillas (GH), 
Gaussian in Age (GIA) and an analytical form popularized by Greisen (Greisen).
As part of this analysis, we have generalized the GIA and Greisen functions
to 4-parameters consistent with the GH function.  To then study the 
relation between the three shower parameterizations, and
to use more physically descriptive parameters, this work 
uses the shower full width half-maximum, $fwhm$, 
and shower asymmetry parameter, $f$,
to parameterize the {\it shape} of the shower profiles.  For
profiles with the same ($fwhm$, $f$), the GH and Greisen shower profiles
are essentially identical and systemically less than GIA for shower
depths away from shower maximum.  Of the three functions, 
GH is most convenient as the integral of the GH profile is an 
analytic function.
Monte Carlo simulated air showers using CONEX, and parameterized in
terms of the new parameters: ($fwhm, f$), have correlations 
(between those parameters) greatly reduced over the standard parameterizations
{\it e.g.} Gaisser-Hillas parameters: ($X_0, \lambda$).   This allows shower 
profile reconstructions to add constraints (if needed) on the mostly 
uncorrelated parameters $fwhm$, $f$.  The CONEX shower simulations suggest
that the shower asymmetry parameter, $f$, may have some sensitivity
to the incident cosmic ray particle type: {\it e.g.} p, C/N/O, Fe or $\gamma$.

\section{Acknowledgments} \label{sec:ackno}
We wish to recognize discussions with Miguel Mostafa, Paul Sommers,
Jaroslav Urbar and Marcelo Vogel
during early work on this paper.  We also recognize and appreciate
support by DOE grant DE-FR02-04ER41300.  



\end{document}